\documentclass[conference]{IEEEtran}
\usepackage{amsmath}
\usepackage{algorithm}
\usepackage{algpseudocode}
\usepackage{array}
\usepackage{subcaption}
\usepackage{booktabs}
\usepackage{makecell}
\usepackage{tabularx}
\usepackage{colortbl}
\usepackage{float}
\usepackage{todonotes}
\usepackage{url}
\usepackage{comment}
\usepackage{placeins}
\usepackage[export]{adjustbox}

\ifCLASSINFOpdf
\else
\fi
\begin{document}
%
\title{MDTP---An Adaptive Multi-Source Data Transfer Protocol}

\author{\IEEEauthorblockN{Sepideh Abdollah}
\IEEEauthorblockA{Tennessee Tech University \\Department of Computer Science\\
Email: sabdollah42@tntech.edu}
\and
\IEEEauthorblockN{Craig Partridge}
\IEEEauthorblockA{Colorado State University\\Department of Computer Science\\
Email: craig.partridge@colostate.edu }
\and
\IEEEauthorblockN{Susmit Shannigrahi\\ Tennessee Tech University \\Department of Computer Science\\}
\IEEEauthorblockA{Email: sshannigrahi@tntech.edu}}


%


\maketitle

\begin{abstract}

Scientific data volume is growing in size, and as a direct result, the need for faster transfers is also increasing. 
The scientific community has sought to leverage parallel transfer methods using multi-threaded and multi-source download models to reduce download times. 
In multi-source transfers, a client downloads data from multiple replicated servers in parallel. Tools such as Aria2 and BitTorrent support such multi-source transfers and have shown improved transfer times. 

In this work, we introduce Multi-Source Data Transfer Protocol, MDTP, which  further improves multi-source transfer performance. MDTP logically divides a file request into smaller chunk requests and distributes the chunk requests across multiple servers. Chunk sizes are adapted based on each server's performance but selected in a way that ensures each round of requests completes around the same time.
We formulate this chunk-size allocation problems as a variant of the bin-packing problem, where adaptive chunking efficiently fills the available capacity ``bins" corresponding to each server.  

Our evaluation shows that MDTP reduces transfer times by 10–22\% compared to Aria2, the fastest alternative. Comparisons with other protocols, such as static chunking and BitTorrent, demonstrate even greater improvements. Additionally, we show that MDTP distributes load proportionally across all available replicas, not just the fastest ones, which improves throughput. Finally, we show MDTP maintains high throughput even when latency increases or bandwidth to the fastest server decreases.



\end{abstract}
\section{Introduction}

Over the past decade, scientific data volume in high-energy particle physics, genomics, and astronomy has grown exponentially \cite{arslan2018low, bouhouch2023dynamic, kettimuthu2018transferring}. Similarly, data center traffic at Facebook, Amazon, Microsoft, and Google has also reached exabytes \cite{jain2020system}. These datasets are often replicated for resiliency and throughput. 



Although data replication is widely employed in big data communities, replication is seldom leveraged to enhance transfer performance, rather as a mean to load-balance or ensure failure resilience. Many tools rely on point-to-point transfers (e.g., wget or similar tools) which are slow and cannot utilize all available replicas concurrently. Numerous studies \cite{ jain2020system, chiba2023network, khurshid2021protocols, kim2014multi, marru2023blaze, yildirim2015application} have pointed out this critical issue in big data transfer: the current methods for transferring big data do not fully utilize all available replicas but focus on faster servers.

To speed up data transfer performance, some tools support parallel or concurrent transfers from multiple replicas.
 
Aria2 \cite{aria2} and BitTorrent \cite{cohen2003incentives} are some examples that can download data from multiple sources. However, these tools are not without their problems. They do not always use all available replicas, tend to utilize a set of faster replicas, and leaves the slower replicas underutilized. Moreover, if a server slows down, it is often ignored in favor of faster replicas \cite{aria2-manual}. This approach creates imbalance in requests since faster replicas receive more requests while slower replicas may not be used at all.

Rodriguez et al. \cite{rodriguez2002dynamic} studied file download by dividing the data into equal-sized chunks and distributing them across multiple servers dynamically. The client initially requests different chunks from each server and, as soon as the client finishes downloading from a server, requests another chunk from it. This process continues until the entire file is downloaded. 
Chunk sizes in this method are small enough to allow fine-grained load balancing and ensure all servers finish around the same time, but large enough to minimize idle time between requests. When number of remaining chunks is less the number of active servers, idle servers (typically the faster ones) request the chunks that are still being downloaded. While this approach ends up downloading some chunks twice, it ensures the overall download speed is at least as fast as the fastest server. Our method extends these ideas \cite{rodriguez2002dynamic} further by making two significant changes. First, we do not use fixed sized chunks and second, we do not request the same chunk multiple times --- each chunk is only requested once.

In this paper, we show that there is a potential to improve data transfer performance by utilizing all replicas, fast and slow. Given the different capacity and throughput of replicas, two challenges arise: (a) how do we utilize all replicas according to their capacity? and (b) how do we make sure a slow replica does not slow down overall transfer speed?

In this paper, we introduce Multi-Source Data Transfer Protocol (MDTP), that solves these problems for large-scale data transfers. 
For intelligently utilizing heterogeneous replicas, we divide the file into multiple chunks and request chunks from multiple servers concurrently. We also dynamically adjust chunk sizes based on the observed download speed of previous chunks. This method allocates requests to each replica based on its capacity. Limited research has been conducted in this area, particularly when replica bandwidth varies \cite{li2013parallel, li2017design}. The most common approach for chunk allocation is to divide the file into equal-sized chunks. However, Li \cite{li2017design} theoretically demonstrated that equal chunk sizes fail to fully utilize the available bandwidth. Our work also shows that utilizing dynamic chunk size produces better throughput than using static chunk sizes. 

To address (b), we observe that the multi-source download problem aligns well with variable sized bin-packing technique. Implementing bin packing requires actively assessing each replica's capacity and assigning different sized chunks to each replica. Additionally, it is important to ensure all concurrent download requests complete close to each another so that the client does not need to wait for a straggler to complete their data transfer. This straggler effect is most pronounced when the last chunk request is allocated to a slow replica --- even though the requests to the fastest server finishes, the download is not completed until the last chunk is downloaded from the slow replica. We avoid this problem by dynamically computing bin sizes every request round proportionally to the server's throughput. This approach ensures all requests finish around the same time as the fastest server and we do not need to wait for a straggler to finish. An added benefit to our approach is adaptability to dynamic conditions. If a server's throughput changes during a transfer, we adjust the subsequent request sizes proportionally to the observed throughput.

To evaluate MDTP's performance, we have implemented a prototype using HTTP byte-range requests. We note that our protocol is generic and can be implemented using other methods or protocols. We compared MDTP with a version using equal-sized chunks as well as well-known publicly available tools such as Aria2 and BitTorrent. Our results demonstrate that MDTP outperforms all compared methods, adapts to dynamic network conditions, and utilizes all available replicas. 
Our experimental evaluation demonstrates that MDTP is much faster than BitTorrent. Even when compared to high-performance download tools such as Aria2, our protocol works much better. 
For example, downloading a 64GB file takes an average 445 seconds with MDTP and 516 seconds with Aria2, which is a 13.7\% improvement. In fact, we observe an improvement between 10\% and 22\% across various file sizes when compared to Aria2. Additionally, MDTP improves replica utilization by 17\% relative to Aria2. We discuss detailed evaluation and results in Section \ref{sec:result}.

\section{Background and Related Work}
\label{sec:related-work}
Several widely used protocols facilitate high-speed data transfer. While some of these protocols can retrieve data from multiple sources, to the best of our knowledge, they do not always fully utilize all available sources, particularly when one or more servers are relatively slow. In this section, we analyze the behavior of these systems and provide foundational context for applying bin packing techniques in data transfer.

\subsection{Multi-source Retrieval}
There have been several tools that provides multi-source retrieval. In this section, we discuss some of these tools.

\textbf{Aria2}: Aria2 is a lightweight, multi-protocol, and multi-source file download utility. It supports various protocols, including HTTP, HTTPS, FTP, BitTorrent, and Metalink\cite{grigorean2017evaluation}. 
It is capable of downloading from multiple sources but does not adapt request sizes based on throughput.


\textbf{BitTorrent}: BitTorrent, a peer-to-peer (P2P) \cite{cohen2003incentives} file distribution protocol, efficiently downloads large files by breaking them into equal chunks and distributing the them among peers. However, when one or more sources slow down, overall throughput can also slow down. 


\textbf{GridFTP/Globus}: GridFTP is an extension of the standard File Transfer Protocol (FTP). It is built atop TCP and enhances data transfer capabilities through the use of parallelism, data striping, and dynamic routing techniques \cite{allcock2005globus, yu2015comparative}. 
This protocol also offers capabilities for the transfer of partial data and automatically adjusts TCP buffer/window sizes to optimize performance. It is mostly utilized for transferring large volumes of data between clusters, remote sources, HPC (High performance computing), or set of computers \cite{allcock2005globus,marru2023blaze}. Globus uses the GridFTP protocol to transfer data.

\textbf{Proprietary Tools}: There are a plethora of data transfer tools, such as IBM Aspera, which utilizes the Fast Adaptive Secure Protocol (FASP) for high-speed file transfers \cite{aspera}. Similarly, Signiant provides accelerated cloud-enabled file transfer solutions tailored for media and enterprise workflows \cite{signiant}. Hightail, formerly known as YouSendIt, offers a cloud-based platform for secure file sharing and collaboration \cite{hightail}. Another alternative, Resilio Connect, leverages peer-to-peer technology to enable scalable and resilient file synchronization across distributed systems \cite{resilio}. As commercial products, the specific algorithms used in these are not publicly disclosed \cite{murata2016quality}.


\textbf{mHTTP}
Kim et al. \cite{kim2014multi} proposed mHTTP, which enables the downloading of different parts of a large file from various sources and paths using a modified socket interface. While this work demonstrates the concept of utilizing multiple sources, it does not adapt to changing network conditions.
 
\subsection{Static Chunk Allocation}
The aforementioned works \cite{kim2014multi, cohen2003incentives} have used static and equal sized chunks for distributing resources across multiple servers. However, as Keqin \cite{li2017design} demonstrated, this approach may not be ideal for concurrent or parallel transfers since it does not allow clients or peers (for peer to peer networks) to adapt to changing server throughput. Further, he mathematically proved that retrieving files with varied chunk allocations from multiple peers can significantly enhance the total system transfer time. While the concept of retrieving information from multiple sources for big data applications is well-established, none of the existing studies have thoroughly examined the performance of dynamic chunk size retrieval for data transfers over HTTP. Alternatively, paper \cite{li2017design} has proposed an algorithmic methods specifically for peer-to-peer networks. 

\subsection{Bin Packing Problem}

In this research, we intent to examine the performance of dynamic chunking retrieval for big data transfer while utilizing multiple data sources concurrently. According to paper \cite{li2017design, rodriguez2002dynamic}, to speed up data transfer, chunks should be downloaded at the same time. Thus, to retrieve data almost at the same time, we were inspired by bin packing problem.
Bin Packing is an optimization problem and
the classic version of bin packing algorithm is as follows:
Given a set of objects, where the size \( S_i\) of the \( i\)th object satisfies \(0 < s_i < 1\), the aim is to pack all objects into the minimum number of unit-sized bins. Each bin can contain any combination of objects as long as their total size does not exceed 1, \cite{cormen2022introduction}.
The paper \cite{kang2003algorithms} proposed a variable-sized bin packing problem, where the goal is to minimize the total cost of the bins used. In this problem, the cost per unit size of each bin does not increase as the bin size increases.
\section{Protocol Goals}
\label{sec:problem-statement}
This section discusses the high-level goals of MDTP. Our work has three primary goals for file downloads: (a) leverage all available replicas (b) adapt request sizes based on observed throughput, and (c) minimize download time.

\subsection{Leveraging all replicas} As we mentioned earlier, scientific datasets are often replicated across various organizations. Scientists (and often data download tools) predominately choose the fastest or most familiar servers for data download. This results in high utilization of faster replicas and under-utilization of slower replicas. 

Our goal is to maximize the utilization of all available replicas while distributing the workload based on observed throughput. Additionally, we aim to establish a single TCP connection and a single HTTP session per data source to mitigate the impact of TCP slow-start and HTTP session establishment delays. This approach not only optimizes data transfer efficiency but also ensures fairness among competing applications.

\subsection{Adapt request sizes based on observed throughput: } Efficient data transfer from multiple data sources over wide area networks requires adapting to the server conditions. It is straightforward to cut off slow servers when the throughput goes below a certain threshold. However, it is not easy to utilize available capacity of the slow server by adjusting the request size. 
    
We do this by adapting the chunk size based on the observed throughput. This chunk size is a variable parameter that can adjust to server conditions, and play a crucial role in the efficiency of data transmission. Requesting large-size chunks can simplify system complexity and improve performance.  
However, a single failure can result in the loss of a large chunk and a high response time. Similarly, requesting smaller chunks is simpler and can reduce response times and retransmissions after data loss, however, servers must deal with an additional overhead \cite{song2020analytical} for handling additional requests. The primary objective of our method is to retrieve data from each server according to each server's condition and bandwidth capacity.

\subsection{Minimize download time}
Downloading chunks from slow servers should not negatively impact the total transfer time, ensuring that all chunks are downloaded without waiting for the slow server to complete its downloads. On the other hand, previous work by Rodriguez et al. \cite{rodriguez2002dynamic} demonstrated that maximum speedup is achieved when servers complete their chunk retrieval at the same time. Therefore, it is necessary to have a protocol that allows slow servers to fetch data approximately for the same time as faster servers. We can do this by adjusting the amount of data requested from the slow server. Our method aims to download from all servers simultaneously, ensuring that chunks from different servers with varying speeds are completed at nearly the same time. Calculating the size of such requests renders well to bin packing problem, which we have utilized in this work.

\section{Design}
\label{sec:design}
In this section, we present our design to address the challenges outlined in the problem statement. Initially, we theoretically demonstrate how using multiple sources reduces transfer time. Then, we show how each replica can be efficiently utilized without negatively impacting transfer performance.

\subsection{Comparison of Queuing Models: Multiple Sources (Model A) vs. Single Source (Model B)}

Simple queuing theory shows us downloading from multiple sources is always faster than downloading from a single source - even when the some sources are slower. Here we compare these two models - a client is downloading a file chunked into multiple segments from multiple sources vs from a single source.


In Model A, where multiple sources are available for concurrent processing, we can model the download time for $X$ chunks using the M/M/c queuing model. In this model:

\begin{itemize}
    \item $\lambda$ represents the arrival rate of requests from the client.
    \item $c$ is the number of servers.
    \item $\mu_i$ represents the service rate of each individual server.
\end{itemize}

The utilization factor ($\rho$) for Model A is given by:
\begin{equation}
\rho = \frac{\lambda}{\sum_{i=1}^{c} \mu_i}
\end{equation}

In Model B, where a single source is used for sequential processing, we can model the download time for $X$ chunks using the M/M/1 queuing model. 
In this model the utilization factor ($\rho$) for Model B is given by:

\begin{equation}
\rho = \frac{\lambda}{\mu}
\end{equation}


\noindent\textbf{Download Times:}
The average download time for a single chunk ($W_A$) in Model A, can be calculated as:

\begin{equation}
W_A = \frac{1}{\sum_{i=1}^{c} \mu_i - \lambda} \quad \text{for } \rho < 1
\end{equation}

The average download time for a single chunk ($W_B$) in Model B, can be calculated as:

\begin{equation}
W_B = \frac{1}{\mu - \lambda} \quad \text{for } \rho < 1
\end{equation}
The key difference between the two models lies in the value of $c$ (the number of servers) and $\sum \mu_i$ (total service capacity). Utilizing multiple servers in parallel means $\sum \mu_i$ is likely to be much higher for model A than $\mu$ in model B under normal circumstances. Since $\lambda$ is constant for both models, the average time to download a chunk in model A is much smaller than the time needed in model B when 
($\rho < 1$) and ($c > 1)$. Given the number of chunks is constant for a file, the total time needed to download the file is also much smaller in model A than model B.

\subsection{Dynamic Chunking and Bin Packing Algorithm}

To achieve the goal of leveraging all available sources and minimizing download time, as outlined in the problem statement, our method draws inspiration from the bin packing algorithm. This requires mapping our model to the bin packing problem. As mentioned in the related work section, the bin packing algorithm involves multiple bins, each with a defined capacity and threshold so that objects must be placed into each bin such that the total value in a bin does not exceed its threshold. To accurately map our problem to this algorithm, we represent each server as a bin, with the server's throughput defining the bin's capacity. The chunk sizes are treated as objects (values) to be distributed among the bins, and the bin's threshold corresponds to the fetching (download) time.

The first step in applying this model is to evaluate each server's throughput (capacity) using an initial uniform chunk request from all of the replicas. This evaluation provides the necessary information to determine the appropriate chunk sizes that each server can handle. However, the key challenge lies in distributing the chunk sizes (values or objects) among servers (bins) with varying capacities. The distribution should be in a way that ensures all chunks are downloaded almost at the same time, maintaining a consistent bin threshold which in this context refers to the download time.

To determine multiple chunk sizes based on server capacities and distribute them among servers (bins), a fixed threshold must be established for all bins. As we mentioned in the related work section, to optimize retrieval time and minimize the impact of slower servers, all servers should complete their chunk downloads at approximately the same time. Since the fastest server has the minimum download time, we consider the fastest server download time as all bins threshold. The fastest server is identified as the one with the highest throughput, calculated by dividing the initial chunk size by its respective download time.

 To select the fastest server among the all server's throughput, we employ the geometric mean rather than using a computationally expensive sorting algorithm. The geometric mean is particularly advantageous because it is less sensitive to outliers, extremely slow servers do not disproportionately affect the overall measure. This characteristic makes it a robust choice for our analysis compared to other inequality measures.

Given \( N \) servers with throughputs (capacity) \( th_1, th_2, \ldots, th_N \), the geometric mean \( \text{GM} \) is calculated as:

\[
\text{GM} = \left( \prod_{i=1}^{N} th_i \right)^{\frac{1}{N}}
\]

If a server's throughput is greater than or equal to the geometric mean, it is considered a fast server; otherwise, it is considered a slow server. Among the fast servers, we identify the fastest one to calculate the download time for each bin. Based on the fastest server's throughput and the initial chunk it downloaded, we can measure its download time, which sets the capacity for all bins. 
If the throughput of the fastest server is \( th_{\text{fastest\_server}} \) and \( C \), the chunk that it downloaded is the initial chunk, the download time for this server is given by:
\[
T_{\text{fastest\_DownloadTime}} = \frac{C}{th_{\text{fastest\_server}}}
\]

The chunk size for each server is proportional to the throughput  of each server \( th_i \), multiplied by the fastest server download time \( T_{\text{fastest\_server}} \). This relationship is mathematically represented as follows:

\[
C_i = T_{\text{fastest\_DownloadTime}} \cdot th_i
\]
Given that \( C_i \) for \( C_1, C_2, \ldots, C_N \) represents the size of each chunk.

\begin{figure}[!ht]
    \centering
    \includegraphics[width=3in]{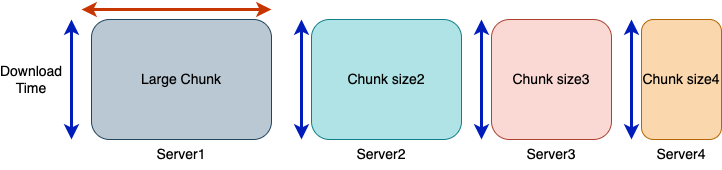}
    \caption{Figure showing variable sized chunks for different servers with different speeds to finish in same time in MDTP}
    \label{Bin packing}
    \vspace{-1em}

\end{figure}

All servers then adjust their chunk sizes to ensure they finish downloading their chunks at nearly the same time. After obtaining the throughput of all servers in each iteration, the fastest server is selected, and the chunk size for all servers is calculated accordingly. As this process is repeated in each round, we ensure that changes in a server throughput, do not affect the download time for each bin in each iteration.
To determine which bin should receive a chunk first, we employ the concept of the best fit algorithm. In our approach, as soon as a server becomes available, it fetches its corresponding chunk that has been pre-calculated by the client. Figure \ref{Bin packing} clearly illustrates the bin packing concept used in our method. In this figure, each server is represented as a bin with varying sizes to reflect their different capacities. The vertical red arrow indicates the differing widths (capacity) of each bin. The blue horizontal arrow represents the download time (threshold), which is the same for all servers. Therefore, within this download time, each server can download objects of different values, where the value in our context refers to the chunk size. As shown in Figure \ref{Bin packing}, the server with higher capacity is assigned a larger chunk, while the server with lower capacity is assigned a smaller chunk. In this method, if all servers have identical capacities, the chunk sizes will be almost equal.

\begin{algorithm}
\caption{Server Chunk Size Calculation}
\begin{algorithmic}[1]
    \State $initial\_chunk\_size \gets C$
    \State $large\_chunk\_size \gets L$
    \State $global\_start\_byte \gets 0$
    \State $global\_end\_byte \gets 0$
    \ForAll {servers $i \in \{1,2, \dots, N\}$} \textbf{in parallel} 
        \State $global\_end\_byte \gets global\_start\_byte + C - 1$
        \State $start\_byte \gets global\_start\_byte$
        \State $end\_byte \gets global\_end\_byte$
        \State $global\_start\_byte \gets global\_start\_byte + global\_end\_byte - start\_byte+ 1$
        \State $server\_i.throughput \gets fetch\_range(start\_byte, end\_byte)$
    
        \While{$global\_end\_byte < file\_size$ \textbf{and} $global\_start\_byte < file\_size$}
    
        \State $product\_of\_throughputs \gets 1$
    
        \State $product\_of\_throughputs \gets product\_of\_throughputs \times server\_i.throughput$
    
        \State $geometric\_mean \gets product\_of\_throughputs^{(1 / \text{N})}$
    
            \If{$server\_i.throughput \geq geometric\_mean$}
                \State $fast\_servers\_throughput\_list \gets server\_i.throughput$
                \State $fastest\_server\_throughput \gets max(fast\_servers\_throughput\_list)$
                \State $fastest\_download\_time \gets L / (fastest\_server\_throughput)$
                \State $large\_start\_byte \gets global\_start\_byte$
                \State $large\_end\_byte \gets large\_start\_byte + L$
                \State $global\_end\_byte \gets large\_end\_byte$
                \State $global\_start\_byte \gets global\_start\_byte + large\_end\_byte - large\_start\_byte + 1$
                \State $fetch( large\_start\_byte, large\_end\_byte)$
                \State 
            \EndIf
                \State $small\_chunk \gets round(fastest\_download\_time \times server\_i.throughput)$
                \State $small\_start\_byte \gets global\_start\_byte$
                \State $small\_end\_byte \gets small\_start\_byte  + small\_chunk$
                \State $global\_end\_byte \gets small\_end\_byte$
                \State $global\_start\_byte \gets global\_start\_byte + small\_end\_byte - small\_start\_byte + 1$
                \State $fetch( small\_start\_byte, small\_end\_byte)$
                
        \EndWhile
    \EndFor
\end{algorithmic}
\end{algorithm}


\section{Implementation}
\label{sec:implementation}
\textit{Note that our prototype implementation of MDTP is in Python and inherently slow. Even though MDTP is implemented in Python, it often outperform Aria2, which is written in C++\cite{aria22025Mar}.}

In this work, we implemented two transfer protocols: (i) MDTP, and (ii) a reference multi-source protocol which utilizes static chunking \cite{rodriguez2002dynamic}. 
For these implementations, we leveraged the aiohttp \cite{aio-libs2023Nov} and asyncio libraries \cite{python_asyncio}
to facilitate asynchronous HTTP requests and responses so that we can handle multiple data transfers concurrently without blocking. 

To ensure fairness and prevent issues like server overload and network congestion, our method establishes only one session per server connection and reuses it.
Chunks are downloaded asynchronously within these sessions. This approach also ensures we do not reset the TCP congestion window at each request round.

We implemented a static chunking-based transfer protocol. Unlike our dynamic chunking method in MDTP, which employs the bin packing algorithm to optimize chunk distribution, this model uses a uniform chunk size across all servers. This is similar to the work done by Rodriguez et al \cite{rodriguez2002dynamic}. 
It shares the core features and operational details of MDTP, with the primary difference being its chunk-sizing strategy. 


\begin{figure*}[!ht]
    \centering
\begin{subfigure}[t]{0.3\textwidth}
  \centering
  \includegraphics[width=\linewidth]{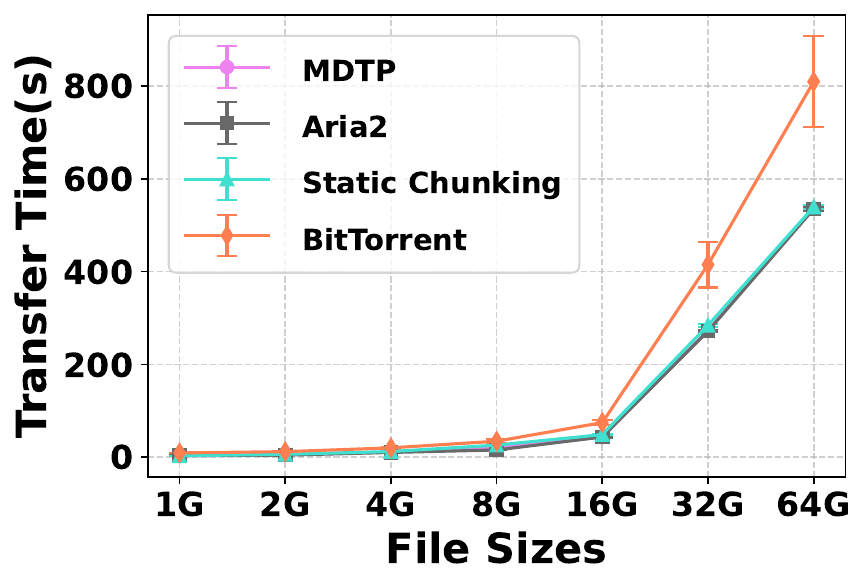}
    \caption{}    
    \label{fig:AllProtocolComparison}
\end{subfigure}  
\begin{subfigure}[t]{0.3\textwidth}    
  \includegraphics[width=\linewidth]{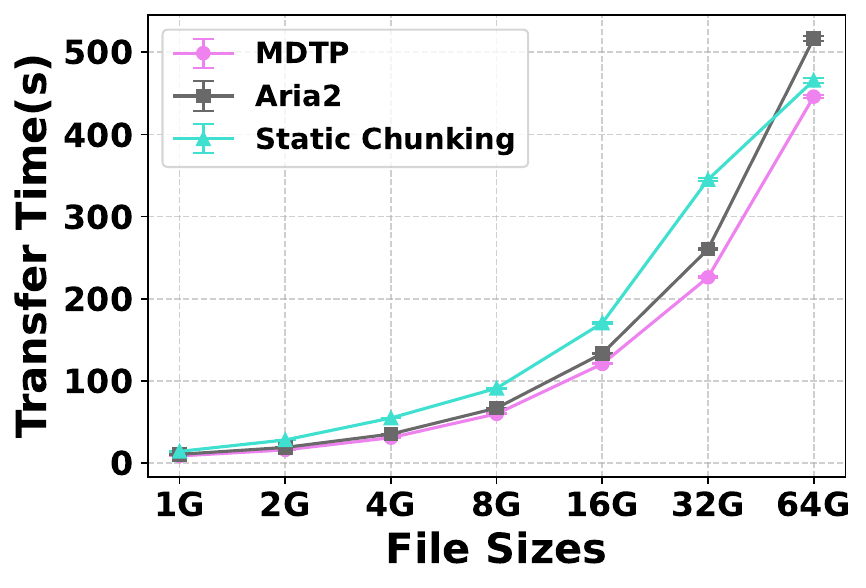}
    \caption{}
        \label{fig:WithoutWritingDisk}
      \end{subfigure}
\begin{subfigure}[t]{0.3\textwidth}
  \centering
  \includegraphics[width=\linewidth]{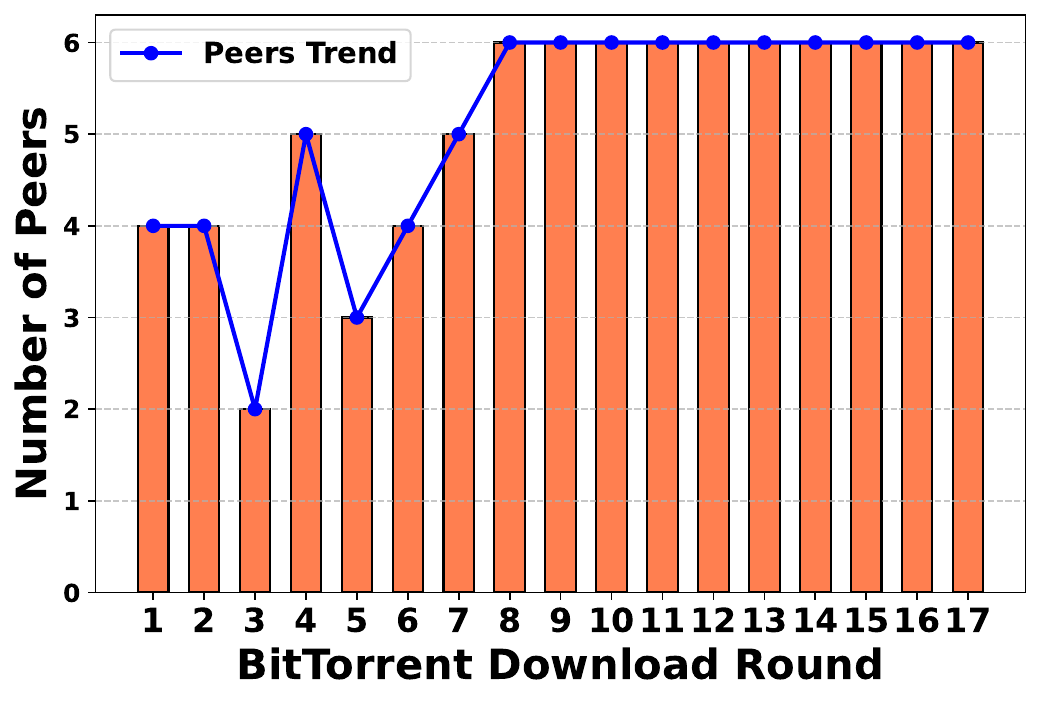}
    \caption{}    
    \label{fig:PeersTrend}
\end{subfigure}
    \caption{ (a) Average transfer time trend including disk write delay. MDTP, Aria2, and Static Chunking perform similarly while BitTorrent is much slower, (b) Average transfer time trends without disk write delay.   We omit BitTorrent due to its lower performance. MDTP performs better than Aria2 and Static Chunking for larger file sizes, and (c) Number of peers seeding a 2GB file in different download iterations. The unpredictable number of seeders make BitTorrent unsuitable for this use case.}

\end{figure*}

\section{Experimental Setup}
\label{sec:experimental-plan}
To assess the effectiveness of MDTP for large file transfers, we created an experimental setup on the FABRIC testbed \cite{fabric-2019} and compared MDTP's performance against BitTorrent, static chunking, and Aria2.  We do not compare our work with GridFTP/Globus since in multi-source scenarios, Aria2 performs better than Globus as demonstrated by \cite{8123734}.

\subsection{File and Chunk Sizes}

We selected file sizes of 1GB, 2GB, 4GB, 8GB, 16GB, 32GB, and 64GB for our experiments. Although scientific data is cumulatively large, this range covers the file sizes found in scientific communities. We did not experiment with very large files since individual file sizes are not huge in scientific communities - the datasets are collection of GB sized files and cumulatively large.

To determine the optimal initial and large chunk sizes for minimizing transfer time in MDTP, we conducted additional experiments using various chunk sizes. These experiments were performed across different file sizes: 1GB, 2GB, 4GB, 8GB, 16GB, 32GB, and 64GB.
Our findings show that for file sizes of 1GB, 2GB, 4GB, and 8GB, the optimal chunk sizes remained nearly the same, with 4MB as the initial chunk size and 40MB as the large chunk size. For files larger than 8GB, the optimal chunk sizes increased, with the initial chunk size rising to 16MB and the large chunk size to 160MB. While we picked these sizes experimentally here, we plan to automate chunk size selection in a future work. 

To determine the optimal chunk size for Static Chunking, we experimented with a range of chunk sizes from kilobytes to megabytes for each file. We then picked the chunk sizes that achieved the minimum transfer time.

\subsection{Sever-Client Deployment for MDTP}
We set up a client and six geographically distributed servers 
on the FABRIC Testbed \cite{fabric-2019}.
We used Apache webserver \cite{ApacheHTTPServer} to serve files of various sizes to the client. The system specifications for both servers and the client are in Table \ref{tab:System_Specification}. We ensured that each server had sufficient capacity, and there were no server-side bottlenecks (unless where we artificially introduced such bottlenecks).


On the client side, MDTP used HTTP/2 range requests to download these files. We ensured there were no client-side bottlenecks so that we could perform a fair evaluation of the system's performance.




To compare MTDP with a well-known, high-speed downloading tool, we utilized Aria2.
Since it supports HTTP, we used the same client-server setup as MDTP. The system specifications were also the same as those in Table \ref{tab:System_Specification}.

We also compared MDTP with BitTorrent due to its widespread use as a distributed file transfer protocol \cite{mazurczyk2013understanding}. We set up an identical environment on FABRIC with six seeders and one client. All seeders had the file and were actively seeding while the client (leecher) just downloaded the files. 
To maximize performance, we configured BitTorrent to utilize its full bandwidth and modified the choking algorithm so that all peers seeded the client without expecting chunks in return.  We used Aria2 \cite{aria2} for seeding and leeching, generated torrent files with Transmission-CLI \cite{transmissioncli}, and set up our own tracker using OpenTracker \cite{opentracker} to manage peer connections.


\begin{table}[h!]
\centering
\resizebox{0.5\textwidth}{!}{
\begin{tabular}{cccccc}
\toprule
\textbf{File Size (GB)} & \textbf{OS} & \textbf{Virtual Machine} & \textbf{RAM (GB)} & \textbf{Disk (GB)} & \textbf{Network Speed} \\
\midrule
1, 2, 4, 8, 16, 32, 64  & Ubuntu  & Fabric Testbed  & 128  & 1000 & 10Gbps\\

\bottomrule
\end{tabular}%
}
\caption{System specification for client and servers  on FABRIC}
\label{tab:System_Specification}
\end{table} 

\section{Evaluation}
\label{sec:result}

For evaluating MDTP, we downloaded different sized files using each download tools and compared their performances.  We conducted five types of experiments to evaluate performance under different network conditions. We repeated each experiment ten times, and averaged the results for accuracy. 
To better assess the differences and accuracy among the ten iterations, we calculated their standard error. A higher standard error indicates greater variation among the data, whereas a lower standard error signifies more consistency.

\begin{figure*}    
\centering
    \begin{subfigure}[t]{0.24\textwidth}
        \includegraphics[width=\linewidth]{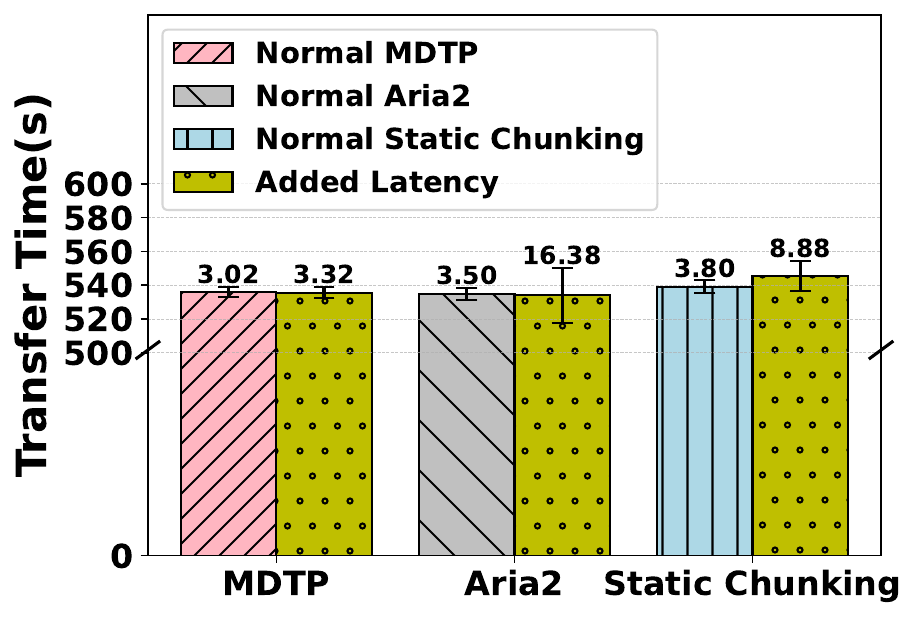}
        \caption{}
        \label{fig:addDelay}
    \end{subfigure}
\begin{subfigure}[t]{0.24\textwidth}
        \includegraphics[width=\linewidth]{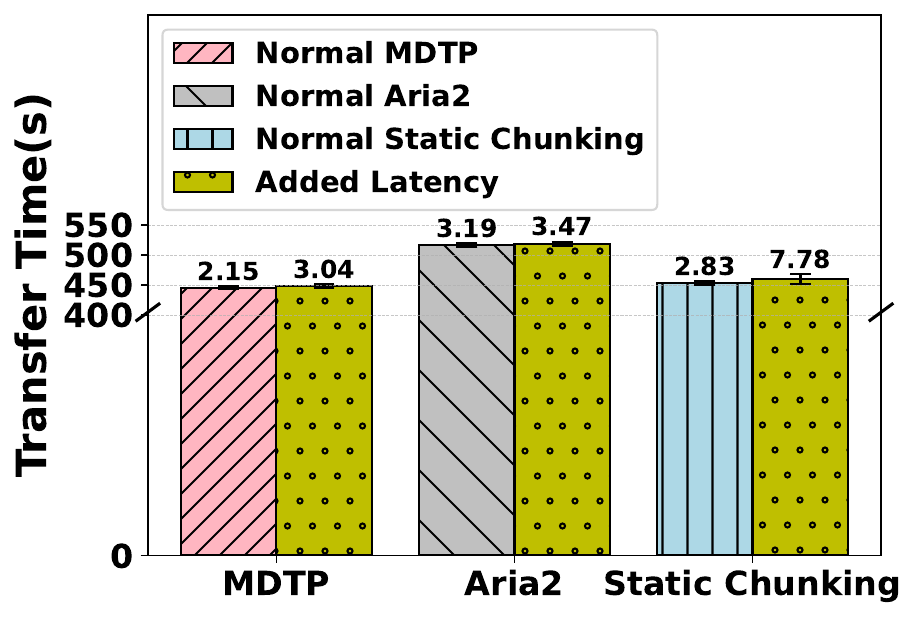}
        \caption{}
        \label{fig:LatencyWithoutDisk}
    \end{subfigure}
    \centering
    \begin{subfigure}[t]{0.24\textwidth}
        \includegraphics[width=\linewidth]{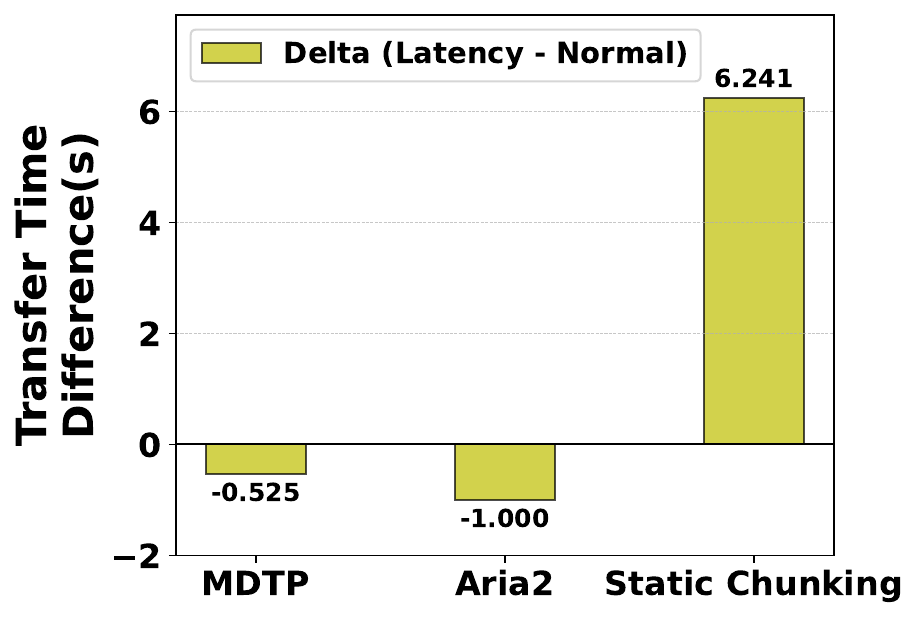}
        \caption{}
        \label{fig:deltaWithoutDisk}
    \end{subfigure}
\begin{subfigure}[t]{0.24\textwidth}
        \includegraphics[width=\linewidth]{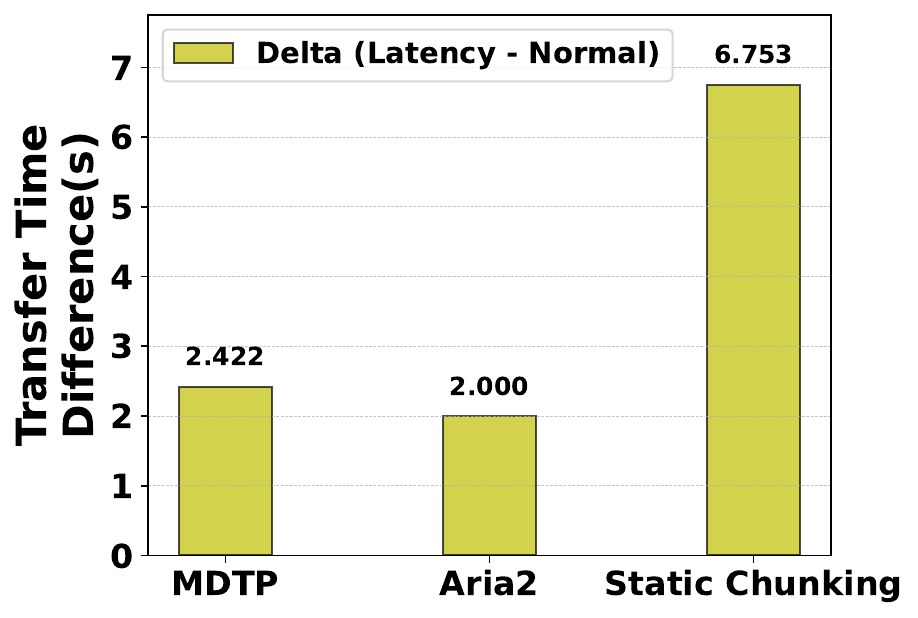}
        \caption{}
        \label{fig:WithoutIODelay0.5sLatency}
    \end{subfigure}
    \caption{(a) Average transfer times with and without 0.5s added latency for a 64GB file, including disk write delay. MDTP and Aria2 exhibit similar behavior, with Aria2 being more inconsistent, while Static Chunking has higher latency. (b) Transfer times with and without 0.5s added latency for a 64GB file, excluding disk write delay. MDTP achieves 13.54\% improvement over Aria2 and 2.6\% over Static Chunking under 0.5s latency. (c) Transfer time delta with and without 0.5s added latency, including disk write delay. MDTP and Aria2 improve by 0.52s and 1s under 0.5s latency, respectively while Static Chunking experiences an extra 6.24s latency. (d) Transfer time delta with and without 0.5s added latency, excluding disk write delay. MDTP and Aria2 had similar latency under 0.5s, while Static Chunking experienced nearly three times higher latency.}
\end{figure*}

\begin{figure}
     \begin{subfigure}[t]{0.24\textwidth}
        \includegraphics[width=\linewidth]{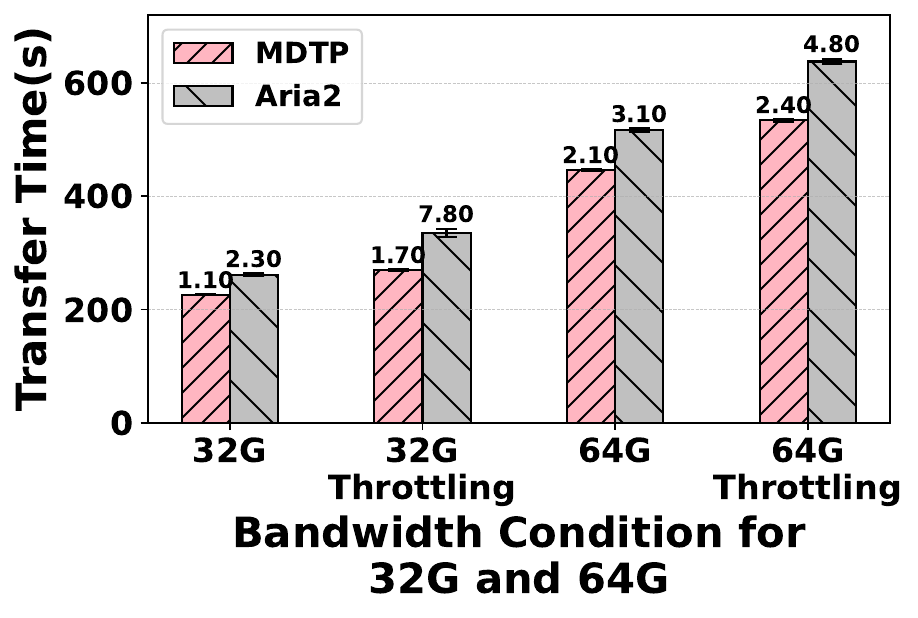}
        \caption{}
        \label{fig:LimitedBW64G}
    \end{subfigure}
    \hfill
    \begin{subfigure}[t]{0.24\textwidth}
        \includegraphics[width=\linewidth]{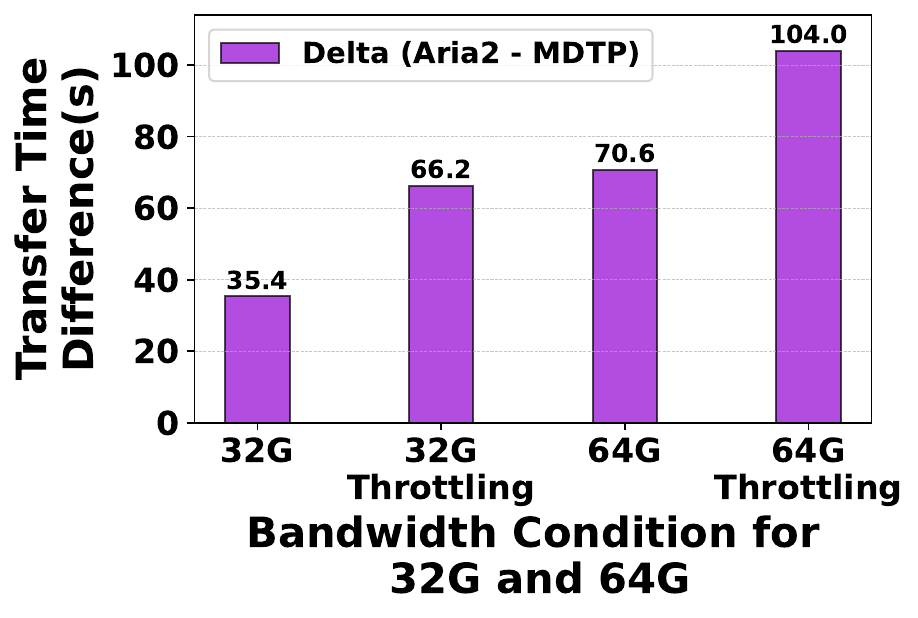}
        \caption{}
        \label{fig:BandwidthDelta}
    \end{subfigure}
    \caption{(a) MDTP vs. Aria2 transfer time for 32GB and 64GB file sizes with and without bandwidth throttling. Bandwidth throttling increases transfer time for both MDTP and Aria2, with a greater impact on Aria2. (b) Transfer time delta of MDTP and Aria2 under different bandwidth conditions. The delta between MDTP and Aria2 increases with bandwidth throttling, as Aria2 does not utilize the slower replicas.}
\end{figure}

\begin{figure*}[!h]
       \begin{subfigure}[t]{0.33\textwidth}
          \centering
        \includegraphics[width=\linewidth]{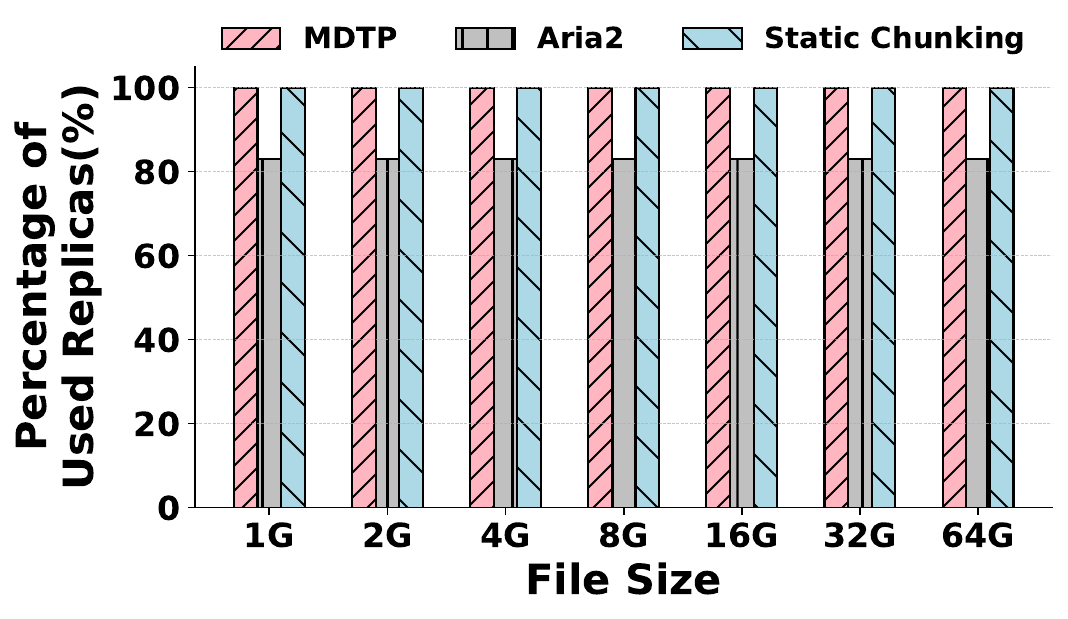}
        \caption{}
        \label{fig:usagePercent}
    \end{subfigure}
    \begin{subfigure}[t]{0.33\textwidth}
        \includegraphics[width=\linewidth]{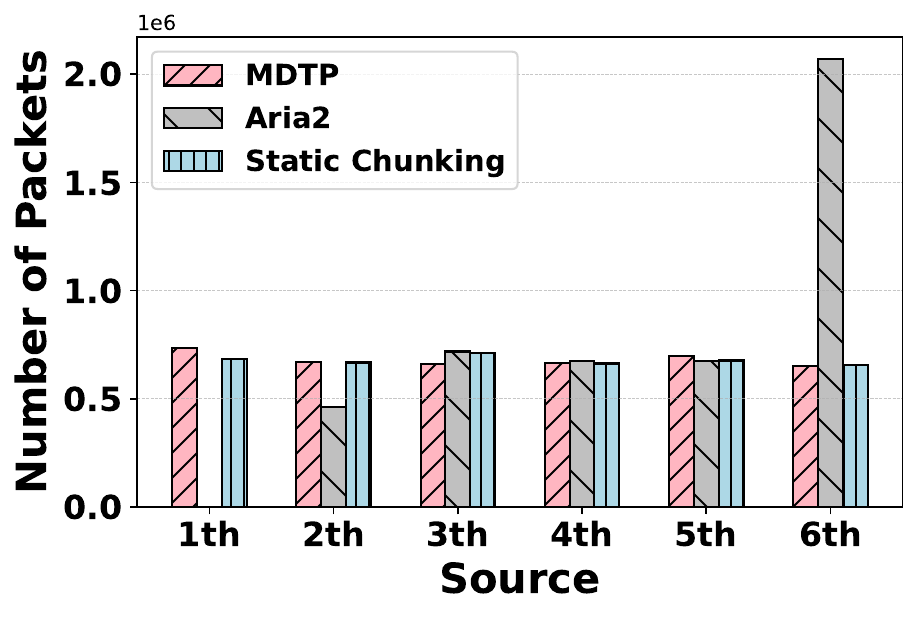}
        \caption{}
        \label{fig:RequestNumber}
    \end{subfigure}
    \begin{subfigure}[t]{0.33\textwidth}
        \centering
        \includegraphics[width=\linewidth]{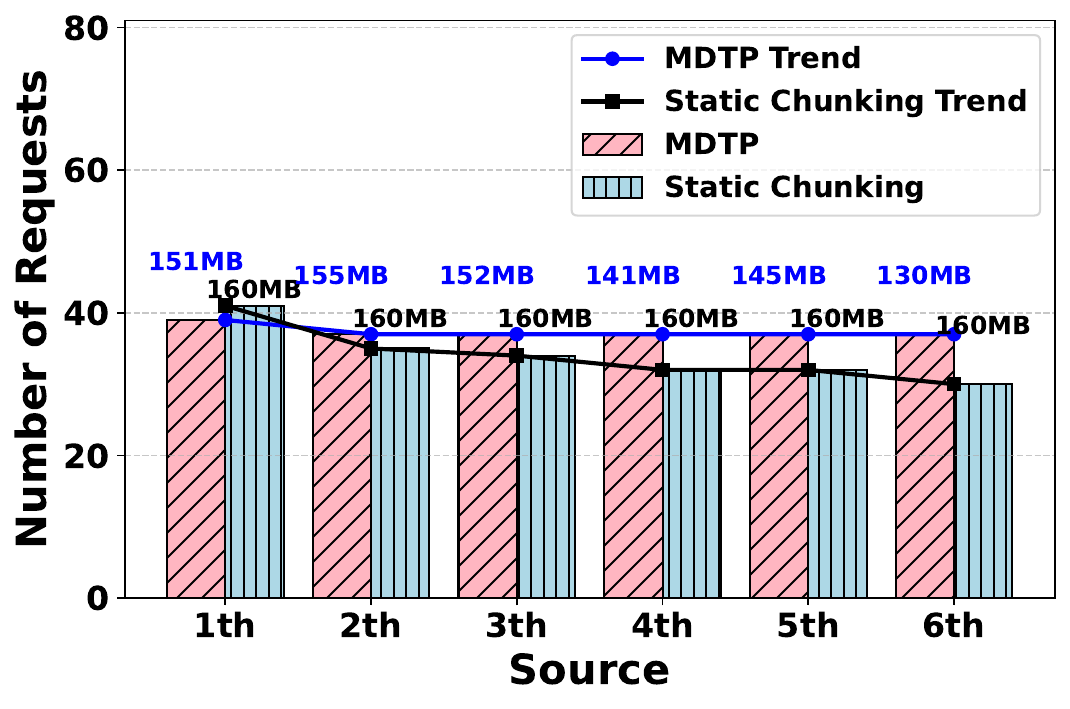}
        \caption{}
        \label{fig:MDTPvsStaticRequests}
    \end{subfigure}
    \caption{(a) Percentage of available replica utilization for different file Sizes. MDTP and Static Chunking utilize all replicas, while Aria2 uses only 83\% of them. (b) Number of packets sent to replicas for a 32GB file transfer. It shows MDTP and Static Chunking are balanced but Aria2 burdens the fastest replica. (c) Number of requests and request sizes in MDTP and Static Chunking for a 32GB File. MDTP demonstrates better load balancing compared to Static Chunking.}
\end{figure*}

\subsection{Experiment under normal network condition}

First, we conducted baseline experiments comparing MDTP, Aria2, Static Chunking, and BitTorrent under normal conditions, without bandwidth limitations or added latency.
We transferred various file sizes (1GB, 2GB, 4GB, 8GB, 16GB, 32GB, and 64GB) using all four protocols.
As Figure \ref{fig:AllProtocolComparison} shows, the performance of MDTP, Aria2, and Static Chunking were similar.
 BitTorrent, however, achieved a significantly lower throughput, with nearly twice the transfer duration of MDTP, Aria2, and Static Chunking obviously for 32GB and 64GB as shown in Figures \ref{fig:AllProtocolComparison}.  
 Additionally, the consistently high standard error across most scenarios highlights BitTorrent's relative instability, with the worst-case standard error reaching 98.27, 32 times higher than MDTP's 3.019.
 For file sizes up to 16GB, we calculated the average transfer time under ideal BitTorrent conditions where  all peers were seeding. 
 
 We found BitTorrent to be unpredictable. We turned off BitTorrent chocking mechanism in the configuration, and yet, not all replicas seeded all the time. In some cases (e.g., 32GB and 64GB), not all six peers participated in seeding. in these cases, we averaged the data transfer time based on scenarios where up to five seeders were actively contributing. Figure \ref{fig:PeersTrend} further illustrates fluctuations in seeder activity. For some rounds, four seeders seeded, then number of active seeders dropped to two, then increased to five, and so on. This shows even when all replicas are up and seeding, not all seeders participare in data delivery. In addition to lower performance, we conclude such unpredictable behavior is not suitable for scientific data transfers. Troubleshooting its erratic seeding behavior proved challenging due to multiple factors, including OpenTracker configuration and individual peer behaviors. Given these inconsistencies, we excluded BitTorrent from subsequent experiments.


\subsection{Experiment with and without disk delay}
Analyzing our code, we identified that disk delay in MDTP is significantly high. This increased delay is attributed to python implementation and our approach of flushing chunks to the disk serially rather than in parallel.

Since the contribution of this work is not developing a high-performance tool but to present the MDTP protocol, we reran the transfer experiments without 
writing data to the disk. We configured MDTP to discard downloaded data and Aria2 to direct its output to /dev/null, thereby bypassing disk write operations. We present the results in Figures \ref{fig:WithoutWritingDisk}.  We find MDTP consistently achieves the lowest transfer time across all file sizes. This figure shows that with an increase in file size, transfer time rises across all protocols, but the gap between MDTP and Aria2 widens significantly at larger file sizes (e.g., 32GB and 64GB). For downloading a 64 GB file, Aria2 took 516.6 seconds while MDTP took 445.9 seconds, a 13.69\% improvement. Additionally, while Aria2 generally outperforms static chunking, its performance declines below static chunking at 64GB, making it less efficient at larger scales.

\subsection{Experiment by added network latency to the fastest server}
Given network conditions and latency fluctuate, we artificially introduced 0.5 second latency to the fastest server. 
We reran the throughput evaluation using the 64GB file and evaluated the performance of three methods: MDTP, Aria2, and Static Chunking. Even after  adding this latency, MDTP slightly outperforms Aria2 when transfer times include disk delay. However, without disk delay, MDTP outperformed Aria2 by at least 50 seconds as Figure \ref{fig:LatencyWithoutDisk} shows. This is a  a huge improvement over Aria2.

More surprisingly, adding latency resulted in reduced download times. Download times went down by 0.52 and 1 second for MDTP and Aria2, respectively, as we can see in Figure \ref{fig:addDelay} and \ref{fig:deltaWithoutDisk}. Without disk delay, adding latency increases transfer time by only 2.42 and 2.0 seconds, respectively. Figure \ref{fig:WithoutIODelay0.5sLatency} shows that the impact of added latency on MDTP and Aria2 is minimal. In some cases, as illustrated in Figure \ref{fig:deltaWithoutDisk}, the added latency can even result in faster transfers.

However, Static Chunking experienced a significantly longer transfer time of 6.241 seconds and 6.75 seconds, as shown in Figure \ref{fig:deltaWithoutDisk} and \ref{fig:WithoutIODelay0.5sLatency}. 

This is an interesting observation. Increasing the latency to the fastest server may decrease download time because both MDTP and Aria2 can direct requests to other servers or adjust request sizes. Static chunking, on the other hand, does worse because the number of requests to the affected server stays the same.


\subsection{Experiment by limiting bandwidth to the fastest server}
To understand the effect of bottlenecks, we limited the bandwidth to the fastest server to 500 Mbps. 
Static Chunking was unable to adapt to these constraints, resulting in excessively long transfer times. As a result, we excluded it from these experiments. 
As Figure \ref{fig:LimitedBW64G} shows, imposing a bandwidth limit increases the transfer time for both MDTP and Aria2. Specifically, with this bottleneck, MDTP took 42 seconds longer for downloading a 32GB file and 48 seconds longer for downloading a 64GB file. With the same restriction, Aria2 took 74 seconds and 121 seconds longer for downloading 32GB ang 64GB files, respectively. Figure \ref{fig:LimitedBW64G} also shows the performance gap between MDTP and Aria2 became more significant under bandwidth limitations compared to unrestricted conditions. As Figure \ref{fig:BandwidthDelta} shows, when the bandwidth is restricted, Aria2 took 1.87 times longer than MDTP for downloading a 32GB file and 1.47 times longer to download a 64GB file. This slowdown is caused by Aria2 not utilizing the slower replicas as we discuss in the next section.



\subsection{Chunk distribution among servers}
We evaluated how effectively MDTP, Aria2, and Static Chunking
utilize available replicas by capturing their network traffic using Tcpdump and analyzing the logs. 

For all file sizes,
we analyzed the percentage of available replicas utilized by each method. As  Figure \ref{fig:usagePercent} shows, MDTP and Static Chunking exhibited similar behavior, as both follow the same logic except for chunk size allocation, leading them to utilize 100\% of available servers. 
However, Aria2 consistently used 83\% or 5 out of 6 available replicas. It excludes slower servers and overloads the fastest one. This observation is important since we built MDTP to utilize all replicas, as we discussed in Section \ref{sec:design}. 

Figure \ref{fig:RequestNumber} shows the number of packets to each replica for different protocols. Aria2 sends a large number of packets to the fastest replica (replica 6) but does not use one at all (replica 1). This behavior is not desirable in a multi-source download scenario. In contrast, both MDTP and static chunking utilizes all replicas.

Finally, as Figure \ref{fig:MDTPvsStaticRequests} shows, MDTP demonstrates a more balanced approach compared to Static Chunking, as it fetched an equal number of 37 requests from Replica 2, Replica 3, Replica 4, Replica 5, and Replica 6. In contrast, the number of requests from replicas in Static Chunking varied. The numbers above the bar chart indicate the chunk size requested from each replica. This figure summarizes the fundamental difference between these protocols - the static chunking protocol keeps the chunk sizes constant but varies the number of requests. MDTP, on the other hand keeps the number of requests balanced across replicas while changing the request size. We show in this work that the second approach works better.

\section{Discussions and Future Work}
\label{sec:discussion}

While we have shown how to improve data transfer performance using multiple replicas, there still more room to optimize throughput further. In this section, we discuss a couple of future directions of our work.

\subsection{Automatic Chunk Selection}
 In this work, we find the optimal chunk sizes empirically. Table \ref{tab:chunk_size_ratios} summarizes the selection process of initial and large chunks. We performed experiments with all combinations of initial chunk sizes and largest chunk sizes. We then picked the sizes that are optimal for each file size. In MDTP, we used 4MB and 40MB chunks for 2--8GB transfers and 16MB and 160MB chunks for 16-64GB chunks. This approach is not clearly scalable for arbitrary file sizes. Our future work will explore how to automatically choose these chunk sizes based on network conditions and file sizes.


\begin{table}[h!]
\centering
\resizebox{0.5\textwidth}{!}{
\begin{tabular}{cccccc}
\toprule
\textbf{Initial Chunk Size} & \textbf{Large Chunk1} & \textbf{Large Chunk2} & \textbf{Large Chunk3} & \textbf{Large Chunk4} \\
\midrule
2  & 20  & 10  & 5   & 2.5 \\
\textbf{4}  & \textbf{40}  & 20  & 10  & 5   \\
8  & 80  & 40  & 20  & 10  \\
\textbf{16} & \textbf{160} & 80  & 40  & 20  \\
\bottomrule
\end{tabular}%
}
\caption{Initial and large chunk sizes are given in MB. Bold values represent the optimal initial and large chunk sizes: 4MB and 40MB chunks for 2–8GB file transfers, and 16MB and 160MB chunks for 16–64GB file transfers.}
\label{tab:chunk_size_ratios}
\end{table}

\subsection{Data Integrity and Retransmission}
We will explore a similar data transfer protocol in networks with errors. Because data in MDTP is chunked, it should be more efficient to examine each chunk as they arrive and ask for retransmission as soon as errors are detected. We may  also be able to  determine chunk sizes based on error probability.

\section{Conclusion}
\label{sec:conclusion}
Given that scientific data are often replicated across multiple servers for redundancy, we proposed MDTP, a multi-source data transfer protocol that dynamically alters request sizes to match server capacity.

We designed and implemented MDTP and evaluated its performance under various file sizes and network conditions. Our work indicates that dynamically selecting request sizes using a bin-packing algorithm enhances server utilization, accelerates data transfers, and adapts to the conditions of each replica. 

To assess its effectiveness, we compared MDTP against well-known transfer protocols such as Aria2 and Bittorrent, and a static chunking version of MDTP. The results demonstrate that MDTP outperforms these approaches in terms of transfer time, adaptability, and load balancing across replicas, often by as much as 50 seconds for a 64G file. This performance gain is substantial considering our unoptimized implementation and shows MDTP can provide a huge performance benefit to  communities whose data sizes are approaching Exabytes.



%
\IEEEpeerreviewmaketitle

\bibliographystyle{unsrt} 
\bibliography{ICC}
\end{document}